\documentclass[aps,prper,reprint,longbibliography,floatfix]{revtex4-2}   

\usepackage[T1]{fontenc}
\usepackage[utf8]{inputenc}
\usepackage{times}
\usepackage{array}
\usepackage{titlesec}

\usepackage{geometry}
\newcolumntype{C}[1]{>{\centering\arraybackslash}m{#1}} 
\newcolumntype{L}[1]{>{\raggedright\arraybackslash}m{#1}} 
\newcolumntype{R}[1]{>{\raggedleft\arraybackslash}m{#1}} 
\usepackage{booktabs}     
\usepackage{tabularx}     
\usepackage{ragged2e}     
\usepackage{caption}      
\usepackage{multirow}
\usepackage{placeins}

\usepackage{amsmath}
\usepackage{amssymb}
\usepackage{graphicx}
\usepackage{tikz}
\usepackage{enumerate}
\usepackage{float}
\usepackage{comment}
\usepackage{indentfirst}
\usepackage{tcolorbox}
\usepackage[normalem]{ulem}  
\usepackage{setspace}        

\usepackage{hyperref}
\hypersetup{colorlinks=true,urlcolor=blue,citecolor=blue,linkcolor=blue}
\begin{document}

\title{Using LLMs to Detect Growth in Computational Thinking in Introductory Physics}

 \author{Sean Savage}
 \affiliation{Department of Physics and Astronomy, Purdue University, West Lafayette, IN, 47907, U.S.A.} 

 \author{Anand Shanker}
 \affiliation{Purdue University, West Lafayette, IN, 47907, U.S.A.} 

 \author{Grace Michlitsch}
 \affiliation{Purdue University, West Lafayette, IN, 47907, U.S.A.} 
 
 \author{N. Sanjay Rebello}
 \affiliation{Dept. of Physics and Astronomy / Dept. of Curriculum and Instruction, Purdue University, West Lafayette, IN, 47907, U.S.A.} 

\begin{abstract}
As computation becomes more central to physics education, creating scalable methods to assess authentic computational thinking (CT) in students remains a critical challenge. While student-written responses capture nuanced reasoning, they are difficult to evaluate at scale. In this study, we investigated the use of Large Language Models (LLMs) to analyze students' written explanations of computational physics problems on a pre- and post- semester survey. By first establishing a human-coded baseline, grounded in CT literature, we identified significant growth in Data Practices and Computational Problem-Solving Practices. When given the same responses, an LLM successfully mirrored the human evaluations and scaled up the detection of these key trends across a large dataset. Notably, both human raters and the LLM struggled to reliably evaluate more complex constructs such as Systems Thinking. Overall, this study demonstrates that LLMs offer a viable method to scale the evaluation of students' CT in large-enrollment physics courses.
\clearpage
\end{abstract}

\maketitle
\section{INTRODUCTION}

Computation has become a fundamental domain of science, recognized by the American Association of Physics Teachers (AAPT) as the ``third pillar" of physics alongside experiment and theory \citep{AAPT2016}. As computation becomes increasingly central to science and engineering, physics educators are tasked not only with teaching programming syntax but also with fostering computational thinking (CT). The goal of integrating computation into introductory physics is to shift students away from superficial programming and toward authentic sensemaking, allowing them to model real-world physical systems using established CT practices \citep{weintrop2016definingcomputationalthinking, shute2017demystifying}. However, determining whether students are actually achieving this integrated understanding presents a significant challenge \citep{shepard2000role}.

Historically, physics education research has relied heavily on multiple-choice instruments to measure conceptual understanding \citep{hestenes1992force, thornton1998assessing, singh2016multiple}. While these assessments are scalable and excel at identifying broad performance trends, they are fundamentally limited in their ability to capture students' mental models and reasoning processes in arriving at their choices \citep{Kuechler_Simkin_2003, Roediger2005, rebello2004effect}. Written explanations provide a much richer window into student cognition \citep{nieswandt2009written}. Written assessments require students to explicitly articulate the connections between variables in a script and the physical laws they represent \citep{mcneill2011supporting}, effectively demonstrating their CT practices. Unfortunately, evaluating these open-ended responses at scale is notoriously resource-intensive \citep{kortemeyer2023}. Additionally, achieving acceptable inter-rater reliability often requires many hours of qualitative coding, forcing researchers and instructors to choose between the scalability of superficial assessments and the richness of qualitative data \citep{Casalino2021, BUTCHER2010489}.

This assessment challenge has been intensified by the increasing availability of generative AI, which can generate functional code instantly and allow students to entirely bypass the cognitive effort needed for algorithmic design \citep{zion2024, becker2023,wang2024examining}. To assess learning, assessments must require deep contextual knowledge, where computation is linked to physical scenarios rather than isolated mathematical abstractions. To evaluate such complex written responses at scale, recent advancements in AI offer promising avenues. 
Recent research has demonstrated the feasibility of using AI to evaluate problem solutions \citep{kortemeyer2025, ssavage2025}, rate students' written work \citep{borse2025}, identify student misconceptions \citep{nlp2024misconceptions}, and assist in generating feedback \citep{lee2026towards, allen2025, hashmi2025, Wan_Chen_2024,latif2023fine}. Building upon this work, this study investigates the viability of Large Language Models (LLMs) as a scalable tool for evaluating open-ended CT assessments. Rather than utilizing LLMs to evaluate code correctness, we deployed a custom-prompted LLM \citep{Chen2023} to categorize how participants frame and reason through computational physics problems. To measure the validity of our methods, we address two primary research questions:

\vspace{0.25em}
\textbf{RQ1:} \textit{How do the CT practices of introductory physics students grow in a computationally intensive physics class?}


\textbf{RQ2:} \textit{To what extent can an LLM reliably measure this growth in students' CT at scale compared to humans?}

\vspace{0.4em}

We propose that LLMs can serve as a scalable mechanism for formative assessment, enabling instructors to efficiently track the integration of computational and physical reasoning.

\section{METHODS}

\subsection{Context and instrument development}
To evaluate computational thinking, we anchored our assessment in the well-established taxonomy developed by Weintrop et al. \citep{weintrop2016definingcomputationalthinking}, which categorizes CT in math and science into four distinct domains: Data Practices (DP), Computational Problem-Solving Practices (CPP), Modeling and Simulation Practices (MSP), and Systems Thinking Practices (STP). In our context, DP involves the collection of data and the manipulation of generated data to create visualizations and meaningful conclusions. CPP focuses on decomposing complex physical problems into logical steps to convert physics concepts into functional code and to create computational abstractions. MSP entails constructing and refining computational models to predict the future state of a system based on its current state. Finally, STP involves defining clear system boundaries, synthesizing micro-level interactions into macro-level behavior, and effectively explaining these intricate system interactions. Evaluating all four domains is critical to capturing a holistic view of a student's computational literacy \citep{odden2019}.

To measure these domains within a physics context, three open-ended survey questions were developed. The instrument was iteratively refined by a group of three experts in physics education and computational science to ensure construct validity. The questions were deliberately designed to require the simultaneous application of physics concepts and specific CT practices. Because Systems Thinking (STP) represents the overarching ability to investigate complex systems as a whole and is thus core to physics problems, it was integrated into all three prompts alongside one or more of the other three foundational practices:

\begin{itemize}
  \setlength{\itemsep}{0pt}
  \setlength{\parskip}{0pt} 
  \setlength{\parsep}{0pt}
    \item \textbf{Q25 (DP \& CPP \& STP):} Students interpret a Python-generated mechanical energy vs. position graph of a mass oscillating on a spring.
    \item \textbf{Q26 (CPP \& STP):} Students trace algorithmic logic in a complex Python script modeling an elastic collision.
    \item \textbf{Q27 (MSP \& STP):} Students propose parameters and boundary conditions to design a computational simulation of a sled sliding down a hill.
\end{itemize}

\subsection{Data collection}
Data were collected from students in an introductory, calculus-based engineering physics course at a large public Midwestern university. The course integrated computation heavily into its pedagogy, requiring students to complete weekly labs that included two or more of the above CT practices. Additionally, both labs and recitations required students to complete and submit their work in Python via Jupyter notebook \citep{Jupyter_project}, which is a conducive environment for students to work on their code \citep{Rule2019Jupyter}. 

Students completed the CT assessment online via the Brightspace Learning Management System, proctored by Respondus Monitor \citep{brightspace} to ensure integrity. The assessment was administered twice: as a pre-test during Week 1, and as a post-test during Week 15 after students had engaged with the integrated computationally focused physics curriculum. The questions discussed here were preceded by 24 multiple-choice CT questions, which will be discussed in a future study. A total of 936 students completed both the pre- and post-surveys.  

\begin{figure}[h] 
    \vspace{0.5em} 
    \centering 
    \setlength{\fboxsep}{0pt} 
    \setlength{\fboxrule}{0.5pt} 
    
    \fbox{\includegraphics[trim=1cm 10cm 3cm 1cm, clip, width=0.9\columnwidth]{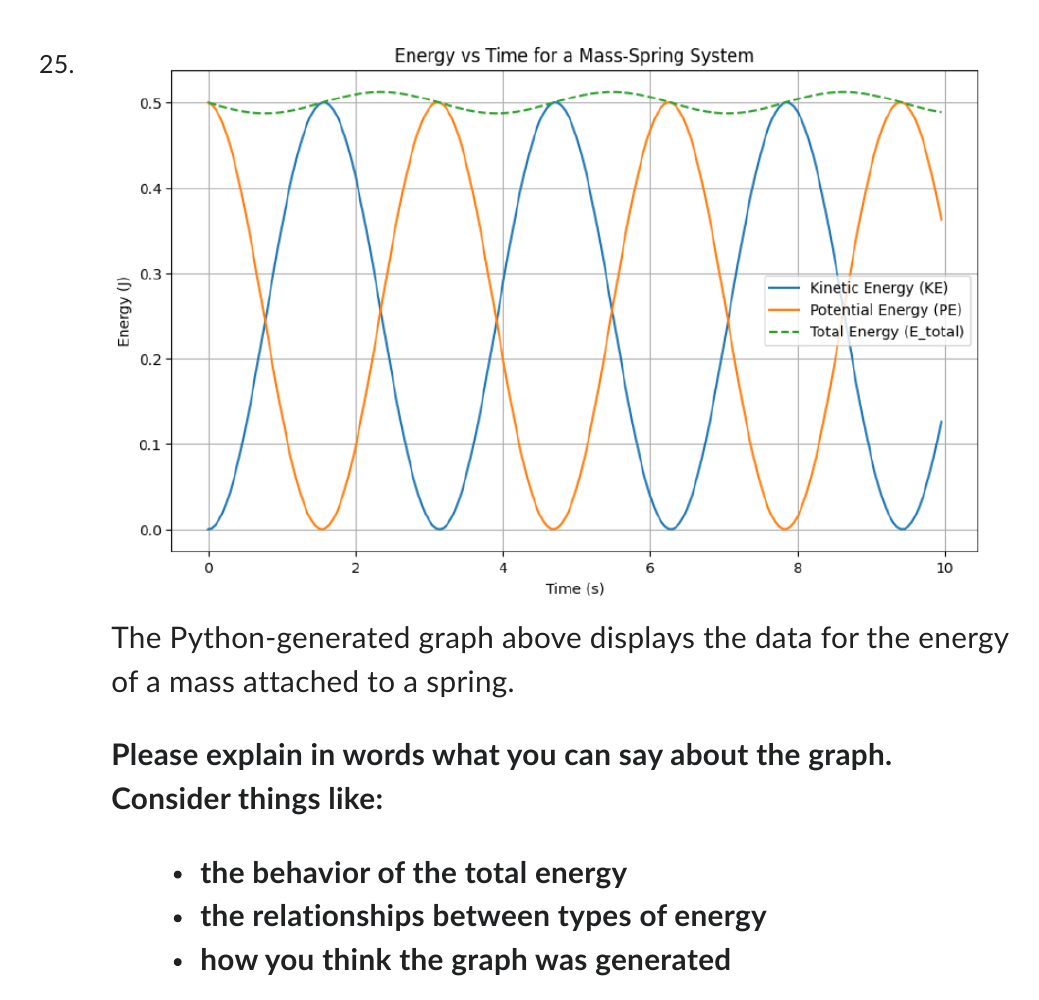}} 
    
    \vspace{0.5em} 
    
    \fbox{\includegraphics[trim=1cm 19.5cm 3cm 6cm, clip, width=0.9\columnwidth]{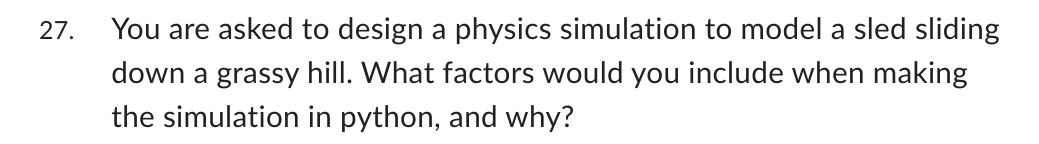}} 
    
    \caption{Two open-ended survey CT questions. Q26 is omitted due to space constraints.} 
    \label{fig:CT_Questions} 
    \vspace{0.25em} 
\end{figure}

\subsection{Data analysis}
Student responses were evaluated using human qualitative coding described below and subsequent LLM validation to track growth in specific CT practices. 

\subsubsection{Human baseline \& rubric development}
To establish a baseline for evaluating student responses, we developed a specialized rubric designed to measure both physical and computational reasoning, based on Wilensky and Weintrop's framework \cite{weintrop2016definingcomputationalthinking}. The rubric was iteratively refined through discussion among the researchers to ensure consistent interpretation. Responses were scored on independent scales for each \textit{CT practice}: a score of 0 represented no evidence, 1 indicated superficial engagement, and 2 indicated deep engagement with the computational concept. For \textit{Physics Correctness}, scores were based on the number of valid physical criteria met: 0 for none, 1 for one criterion, and 2 for multiple criteria. These dimensions were evaluated independently, such that a response could demonstrate valid computational thinking (e.g., noting data overlap on a graph for Data Practices) even if their underlying physical conclusion was incorrect. 

This identical scoring scheme was applied to both human coders and the LLM. Three researchers independently coded an initial subset of $N = 10$ students' pre- and post-instruction responses (60 responses), achieving an average Fleiss' $\kappa = 0.53$ for CT practices and $\kappa = 0.56$ for Physics Correctness \citep{Fleiss1971}. To resolve discrepancies, the researchers reached consensus through iterative discussion, refining the operational definitions of ``superficial'' and ``deep'' engagement. Following calibration, the researchers coded a larger sample of $N = 50$ students (300 total responses), establishing the final human ground truth via majority vote.

\subsubsection{LLM validation \& analysis}
To determine whether generative AI could scale this qualitative evaluation, we used the GPT-5.4-mini API \citep{openai2025chatgpt}. Because the survey questions rely heavily on visual context (e.g., Python code snippets and system diagrams), GPT-5.4-mini was selected for its advanced multimodal capabilities. The model's context window was fed the exact image of the survey question, the prompt text, and the student's written response with each pass. To keep the scoring consistent and allow room for LLM justification, the API temperature parameter was set to $0.2$. The LLM was guided using a structured prompt that incorporated role-playing (instructing the model to act as an expert physics education researcher) and structured reasoning prompts \citep{Chen2023}. The same rubric used by human raters, including scoring criteria, scale definitions, and evidence required for each score level, was applied directly to the LLM through structured prompting, ensuring that both human and LLM evaluations were based on identical standards.
The LLM was prompted to generate structured justifications for specific physics concepts and CT practices before assigning a final 0–2 score. These outputs were first compared against the 50-student human baseline to calculate inter-rater agreement. Once validated against this human consensus, the multimodal LLM pipeline was deployed across the full dataset of $N = 936$ students. To mitigate model inconsistency, every response was evaluated across three independent runs of the LLM, with the final score determined by majority vote. This voting framework perfectly mirrors our human scoring methodology and aligns with established best practices for reducing LLM assessment errors \citep{kortemeyer2023}.
\vspace{1em}

\section{FINDINGS \& DISCUSSION}

\subsection{Human validation} To validate the accuracy of our rubric, a human-coded baseline was first established. Three human researchers independently evaluated a subset of 50 student responses. Inter-rater reliability (IRR) was measured using Fleiss ($\kappa$) \citep{Fleiss1971}. The human consensus demonstrated substantial reliability on explicit computational tasks. Specifically, Modeling and Simulation Practices ($\kappa = 0.80$) and Data Practices ($\kappa = 0.80$) achieved near-perfect agreement. Physics Correctness ($\kappa = 0.73$) and Computational Problem-Solving ($\kappa = 0.60$) demonstrated substantial reliability, while Systems Thinking Practices yielded the lowest agreement ($\kappa = 0.51$), though still within the acceptable range \citep{landis1977measurement}. Because STP requires interpreting a student’s conceptual framing of interacting system components, it is inherently the most cognitively demanding construct to code. As such, this human baseline provides sufficient reliability to support the rubric’s use and establish a reference point for validating the LLM.

\subsection{Student growth in CT}
A paired t-test was conducted on the human-graded sample ($N=50$) to measure overall growth for pre- versus post-instruction student responses (Table \ref{Table_HumanGrowth}). For metrics evaluated across multiple survey items (e.g., Physics Correctness and STP), scores were averaged per student across the relevant prompts before statistical testing. This approach maintained a consistent effective sample size of $N=50$ independent students (with $49$ degrees of freedom) for all comparisons. To evaluate the magnitude of these shifts, Cohen's $d$ effect sizes were calculated alongside standard deviations. Students demonstrated highly significant, large-magnitude growth in Data Practices ($p < 0.001$, $d = 1.03$), as well as moderate growth in Computational Problem-Solving ($p < 0.001$, $d = 0.52$) and Systems Thinking ($p < 0.001$, $d = 0.45$). This substantial improvement in DP aligns with the instructional design of the course, as the weekly laboratory sessions explicitly required students to integrate physical data collection with computational analysis. Overall Physics Correctness also showed a statistically significant, small-to-moderate improvement ($p = 0.008$, $d = 0.31$). Because all significant metrics yielded $p$-values below the strict Bonferroni-corrected threshold of $\alpha = 0.01$ (correcting for five independent comparisons), no further multiplicity adjustments were required. 
However, Modeling and Simulation Practices showed no statistical growth ($p = 1.000$, $d = 0.00$). This lack of change reflects an assessment ceiling effect, in which students entered the course already able to list simulation parameters, leaving no room for measurable growth on this specific prompt.

Analyzing the questions individually revealed a clear compartmentalization of skills (Table \ref{Table_QuestionLevelGrowth}). On Q25, where students reverse-engineered a graph, DP and Correctness grew in tandem. Alternatively, on Q26, students showed highly significant growth in CPP (specifically procedural code tracing), while physics correctness only improved slightly. This is consistent with prior research \citep{aiken2013i}, which shows that students often consider procedural coding as a syntax exercise. For instance, getting better at following loops does not guarantee better physical reasoning unless the prompt explicitly demands physical sensemaking. A similar instance occurred on Q27, where a stagnation in MSP corresponded to a slight drop in physics correctness. This indicates that students' prior knowledge of modeling and simulation saturated their written responses early on. As such, rather than measuring new cognitive growth, their performance on this prompt served primarily as an indicator of the baseline knowledge students brought into the course.

\begin{table*}[t] 
\centering
\captionsetup{justification=raggedright,singlelinecheck=false}
\caption{\small Paired t-test results measuring overall growth in physics correctness and specific Computational Thinking (CT) practices ($N= 50$) performed by human coders. }
\begin{ruledtabular}
\small
\begin{tabular}{lccccc}
\textbf{Metric} & \textbf{Pre Mean $\pm$ SD} & \textbf{Post Mean $\pm$ SD} & $\mathbf{t(49)}$ & \textbf{\textit{p}-value} & \textbf{Cohen's \textit{d}} \\
\hline
Physics Correctness & $1.25 \pm 0.62$ & $1.43 \pm 0.54$ & 2.68 & $<0.05$ & 0.31 \\
Data Practices (DP) & $1.12 \pm 0.66$ & $1.72 \pm 0.50$ & 5.14 & $<0.001$ & 1.03 \\
Comp. Prob-Solving (CPP) & $0.71 \pm 0.77$ & $1.12 \pm 0.79$ & 3.71 & $<0.001$ & 0.52 \\
Systems Thinking (STP) & $0.90 \pm 0.63$ & $1.21 \pm 0.73$ & 3.90 & $<0.001$ & 0.45 \\
Modeling \& Sim (MSP) & $1.62 \pm 0.49$ & $1.62 \pm 0.53$ & 0.00 & 1.000 & 0.00 \\
\end{tabular}
\end{ruledtabular}
\label{Table_HumanGrowth}
\end{table*}
\vspace{0.5em}

\begin{table*}[htbp]
\centering
\captionsetup{justification=raggedright,singlelinecheck=false}
\caption{\small Question-level comparison of pre- and post-instruction means and growth ($\Delta$) across human-graded and AI-graded datasets.}
\begin{ruledtabular}
\small
\renewcommand{\arraystretch}{1.2}
\begin{tabular}{l ccc ccc ccc}
\multirow{2}{*}{\textbf{Question \& Metric}} & \multicolumn{3}{c}{\textbf{Human ($N=50$)}} & \multicolumn{3}{c}{\textbf{LLM ($N=50$)}} & \multicolumn{3}{c}{\textbf{LLM ($N=936$)}} \\
\cline{2-4} \cline{5-7} \cline{8-10}
 & \textbf{Pre} & \textbf{Post} & \textbf{$\Delta$} & \textbf{Pre} & \textbf{Post} & \textbf{$\Delta$} & \textbf{Pre} & \textbf{Post} & \textbf{$\Delta$} \\
\hline
\textbf{Q25: Graph Interpretation} & & & & & & & & & \\
\hspace{3mm} Physics Correctness & 1.14 & 1.56 & +0.42 & 1.04 & 1.54 & +0.50 & 1.05 & 1.48 & +0.43 \\
\hspace{3mm} Data Practices (DP) & 1.12 & 1.72 & +0.60 & 1.24 & 1.82 & +0.58 & 1.25 & 1.74 & +0.49 \\
\hspace{3mm} Comp. Prob.-Solving (CPP) & 0.26 & 0.58 & +0.32 & 0.64 & 0.86 & +0.22 & 0.59 & 0.83 & +0.24 \\
\hspace{3mm} Systems Thinking (STP) & 0.76 & 1.36 & +0.60 & 0.84 & 1.18 & +0.34 & 0.90 & 1.27 & +0.37 \\
\midrule
\textbf{Q26: Code Tracing} & & & & & & & & & \\
\hspace{3mm} Physics Correctness & 1.10 & 1.26 & +0.16 & 1.02 & 1.28 & +0.26 & 1.08 & 1.36 & +0.28 \\
\hspace{3mm} Comp. Prob.-Solving (CPP) & 1.16 & 1.66 & +0.50 & 1.12 & 1.74 & +0.62 & 1.12 & 1.73 & +0.61 \\
\hspace{3mm} Systems Thinking (STP) & 0.76 & 1.12 & +0.36 & 0.83 & 1.16 & +0.33 & 0.83 & 1.11 & +0.28 \\
\midrule
\textbf{Q27: Simulation Design} & & & & & & & & & \\
\hspace{3mm} Physics Correctness & 1.50 & 1.46 & -0.04 & 1.86 & 1.70 & -0.16 & 1.76 & 1.64 & -0.12 \\
\hspace{3mm} Modeling \& Sim. (MSP) & 1.62 & 1.62 &  0.00 & 1.12 & 1.18 & +0.06 & 1.19 & 1.29 & +0.10 \\
\hspace{3mm} Systems Thinking (STP) & 1.18 & 1.14 & -0.04 & 1.22 & 1.22 &  0.00 & 1.22 & 1.11 & -0.11 \\
\end{tabular}
\end{ruledtabular}
\label{Table_QuestionLevelGrowth}
\vspace{1em}
\end{table*}

\subsection{LLM analysis}
To evaluate the performance of the LLM, we first compared its scoring to the same subset of 50 students used for the human baseline. The model achieved substantial agreement and high raw accuracy on explicitly defined practices, including DP ($\kappa = 0.90$, with $93\%$ agreement), MSP ($\kappa = 0.78$, with $85\%$ agreement), and CPP ($\kappa = 0.69$, with $74\%$ agreement). Agreement was lower for more complex, multi-component constructs such as Physics Correctness ($\kappa = 0.53$, with $70\%$ agreement) and STP ($\kappa = 0.48$, with $67\%$ agreement). Unlike more discrete practices, these constructs require raters to identify multiple relevant elements within a response and evaluate their sufficiency and integration, which introduces greater variability in interpretation. These values of Cohen's kappa are still considered to indicate moderate to substantial agreement in the literature \citep{mchugh2012interrater}. Given that the LLM demonstrated comparable patterns of agreement to human raters, particularly for well-defined practices, we next applied the model to the full dataset.

When applied to the full dataset of $N=936$ students, the LLM analysis confirmed the macroscopic trends observed in the human baseline. By evaluating over 2,800 total responses, the model detected highly statistically significant growth ($p < 0.001$) across Q25 and Q26. However, Q27 demonstrated no measurable growth across any parameter, including STP and Physics Correctness. The pre-instruction mean for MSP on Q27 was $1.62$. Because the rubric requires students to identify multiple valid parameters to earn a maximum score of $2$, and the vast majority of students entered the course already possessing the conceptual framing to do so \citep{hammer2000student}, $1.62$ effectively serves as an assessment ceiling. There was mathematically little room left for population-level growth. Consequently, Q27 serves poorly as a pre- versus post-instruction survey instrument. Future iterations of this assessment must revise Q27 to demand deeper integration. For instance, asking students to explain how parameters interact rather than merely asking what factors to include.

\section{CONCLUSIONS, LIMITATIONS, \& FUTURE WORK} 
In response to \textbf{RQ1}, we found that students showed significant growth in Data Practices, Computational Problem-Solving Practices, and Systems Thinking Practices, but little growth in Modeling and Simulation Practices, likely because many entered the course with prior knowledge of MSP. Further, STP was especially difficult to score reliably because it overlapped with other CT practices and required more complex conceptual coding.

In response to \textbf{RQ2}, we found that the LLM approximated human-level evaluation for DP, CPP, and MSP and, when applied to a larger dataset, reproduced similar growth trends. Importantly, the LLM's lower agreement on STP mirrors the lower IRR observed among human coders. This indicates that the LLM performs comparably to human experts, and that the variance stems from the inherent complexity of evaluating the construct rather than a deficiency in the model. Thus, while LLMs provide scalable assessment of CT growth, evaluating highly integrated constructs like STP requires more explicitly operationalized rubrics to achieve high reliability.

As computation becomes increasingly important, we need scalable assessment methods to assess students' CT. We show that an LLM can approximate human coding of students’ written responses, especially for clearly defined CT practices. Agreement was strongest for DP and MSP and moderate for STP, which requires integrating multiple ideas within a response. At scale, the LLM reproduced key human-coded trends, including significant gains in DP and CPP. 

Several limitations remain. Both human raters and the LLM showed lower agreement on multi-component constructs, highlighting the difficulty of assessing complex reasoning in brief written responses. Future rubrics should define STP more explicitly, distinguishing between identifying system components and explaining their interactions and consequences. Some prompts may also lack sensitivity to instructional growth owing to a ceiling effect, due to students' high prior knowledge. Future versions should ask students to explain how parameters interact within the system. Overall, LLMs have strong potential for scalable CT assessment in large-enrollment physics courses, but they require careful rubric design and human oversight for complex reasoning.
\vspace{-1em}
\section{ACKNOWLEDGMENTS}
Supported in part by U.S. National Science Foundation Grants 2300645 and 2111138. Opinions expressed are those of the authors and not the Foundation.
\clearpage

\bibliography{references}
\end{document}